\documentclass[showpacs,twocolumn,floatfix,amsmath,amssymb,superscriptaddress,prl]{revtex4-1}
\usepackage{epsfig}
\usepackage{graphicx}
\usepackage{longtable}

\begin{document}

\title{Odd-frequency triplet pairing in mixed-parity superconductors}
\author{P. Gentile}
\affiliation{Dipartimento di Fisica ``E. R. Caianiello'',
Universit\`a di Salerno I-84084 Fisciano (Salerno), Italy and
CNR-SPIN, I-84084 Fisciano (Salerno), Italy}

\author{C. Noce}
\affiliation{Dipartimento di Fisica ``E. R. Caianiello'',
Universit\`a di Salerno I-84084 Fisciano (Salerno), Italy and
CNR-SPIN, I-84084 Fisciano (Salerno), Italy}

\author{A. Romano}
\affiliation{Dipartimento di Fisica ``E. R. Caianiello'',
Universit\`a di Salerno I-84084 Fisciano (Salerno), Italy and
CNR-SPIN, I-84084 Fisciano (Salerno), Italy}

\author{G. Annunziata}
\affiliation{Dipartimento di Fisica ``E. R. Caianiello'',
Universit\`a di Salerno I-84084 Fisciano (Salerno), Italy and
CNR-SPIN, I-84084 Fisciano (Salerno), Italy}

\author{J. Linder}
\affiliation{Department of Physics, Norwegian University of
Science and Technology, N-7491 Trondheim, Norway}

\author{M. Cuoco}
\affiliation{Dipartimento di Fisica ``E. R. Caianiello'',
Universit\`a di Salerno I-84084 Fisciano (Salerno), Italy and
CNR-SPIN, I-84084 Fisciano (Salerno), Italy}
\date{\today}

\begin{abstract}
We show that mixed-parity superconductors may exhibit equal-spin
pair correlations that are odd-in-time and can be tuned by means
of an applied field. The direction and the amplitude of the pair
correlator in the spin space turn out to be strongly dependent on
the symmetry of the order parameter, and thus provide a tool to
identify different types of singlet-triplet mixed configurations.
We find that odd-in-time spin-polarized pair correlations can be
generated without magnetic inhomogeneities in
superconducting/ferromagnetic hybrids when parity mixing is
induced at the interface.
\end{abstract}

\maketitle {\it Introduction} Symmetry breaking is a central
concept in physics, superconductivity being an exemplary case. In
a conventional superconductor gauge symmetry is broken and Cooper
pairs condense in the most symmetric configuration. The discovery
of superfluid $^3$He started the era of unconventional
superconductivity where other symmetries get broken and
alternative pairing mechanisms emerge with respect to the phonon
mediated one. This happens in high-temperature cuprates,
f-electrons superconductors, cobaltates, ruthenates, pnictides,
organics and many other superconductors
\cite{Norman2011,Pfleiderer2009}.

The issue of the gap symmetry and the pairing mechanism has become
more intriguing since the recent discovery of superconductivity in
a number of materials whose lattices lack inversion symmetry
\cite{Bauer2004}, with important implications for the symmetry of
the superconducting state. In this framework, difficulties may
emerge in detecting parity mixing of Cooper pairs partly due to
the fact that conventional experimental approaches utilized for
the determination of parity of Cooper pairs, such as Knight shift
measurements, do not provide useful information when spin-orbit
coupling is larger than the pairing energy scale.

In some of the unconventional superconductors time reversal
symmetry can also be broken both for spin singlets and spin
triplets as well as in the cases of superconductivity coexisting
with ferromagnetic long-range order \cite{Saxena2000}. Concerning
this symmetry, in the majority of the superconductors the order
parameter is assumed to be even in the frequency domain, so that
it may be even or odd in space depending on whether the Cooper
pairs form spin singlets or triplets, respectively. However, if we
allow for an odd-in-time dependence, more exotic types of pairings
are possible, like that originally predicted in the context of
liquid $^3$He \cite{Berezinskii1974}. For these states the triplet
pair correlation function is symmetric under the exchange of
spatial and spin coordinates but antisymmetric under the exchange
of time coordinates.

A significant advance in the understanding of odd-in-time pairing
has been achieved since many experimental observations
\cite{Nature2006,Sonin2006,Robinson2010,Khaire2010} have been
accounted for as a manifestation of odd-in-time equal-spin (OTES)
pair correlations generated at the interface between a spin
singlet superconductor and a ferromagnet whose magnetization has
an inhomogeneous profile both in amplitude and phase
\cite{Bergeret2001,Bergeret2003}. The interplay of magnetic
non-collinearity and odd-in time pairing emerges also in
magnetically active interfaces that are able to change the spin
direction of incident electrons across the junction
\cite{Eschrig2011,Linder2009}. Otherwise, zero-spin projection
odd-in-time pair correlations take place in normal/superconductor
junctions between triplet superconductors and diffusive normal
metals \cite{Tanaka2007S-DN} or ferromagnets \cite{Annunziata2011}
as well as in conventional ballistic hybrids without spin-triplet
ordering \cite{Tanaka2007S-BN}.

The main interest in the search for odd-in-time pair correlations
is motivated by the possibility of engineering spin-polarized
supercurrents travelling through strong ferromagnets on distances
that do not depend on the strength of the exchange field
\cite{Eschrig2011}. The issue of functionalizing spin-polarized
pair currents in superconducting/ferromagnetic hybrids is indeed a
challenging task for the high potential towards innovative
applications in the field of the spintronics \cite{Zutic04} where
it is the spin degree of freedom rather than the charge one to be
exploited.

The aim of this letter is to demonstrate how odd-in-time
equal-spin pair correlations are intimately connected to the
symmetry breaking in unconventional superconductors. We show that
OTES pair correlations manifest in mixed-parity superconductors
and their existence is strongly dependent on the direction of an
applied field in such a way to single out different types of
parity mixing configurations. We also find a significant
connection with the generation of spin-polarized pairs in hybrids.
A different mechanism, with respect to those mentioned above, is
presented for the case of a superconductor/Rashba
metal/ferromagnet hybrid without requiring any inhomogeneity in
the magnetic profile. Indeed, if parity mixing is yielded at the
interface, i.e. via antisymmetric spin-orbit interaction, then
OTES correlations are not trivial in all directions of the spin
space assuming that the magnetization and the vector defining the
spin-orbit interaction are not orthogonal.

\begin{figure}[bh]
\centering{
\includegraphics[width=0.4\textwidth]{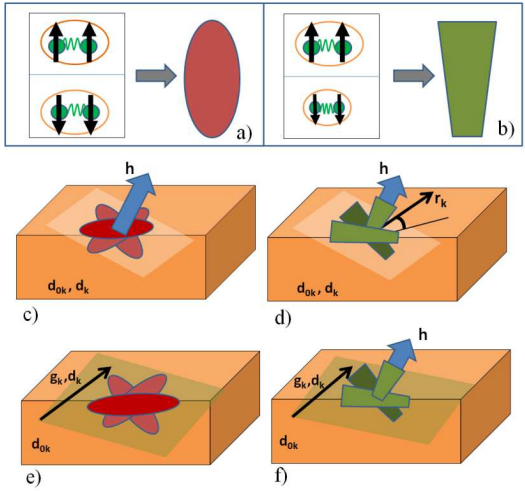}}
\caption{(Color online) Schematic representation of the non-zero
components of the odd-in-time equal-spin pair correlation function
for different types of mixed parity configurations. a) and b)
stand for the symmetric (oval) and asymmetric (trapezoid) spectral
components for the two spin projections of the odd-in-time
equal-spin correlations along a given direction. c) and d)
indicate direction and amplitude of the odd-in-time equal-spin
correlations for a centrosymmetric mixed parity unitary and
non-unitary superconductor with respect to an applied magnetic
field, respectively. e) and f) depict direction and amplitude of
the odd-in-time equal-spin correlations for a non-centrosymmetric
mixed parity without and with and applied field, respectively. The
light rectangle indicates the plane perpendicular to $\textbf{h}$
in c) and d), while it is the plane where
$\textbf{g}_{\textbf{k}}$ and $\textbf{d}_{\textbf{k}}$ lie in e)
and f). See the main text for the definition of
$\textbf{r}_{\textbf{k}}$,$\textbf{g}_{\textbf{k}}$,$d_{0\textbf{k}}$,
and $\textbf{d}_\textbf{k}$.} \label{fig:fig1}
\end{figure}

{\it Model and results} We start by considering a generic
superconducting system having an equal time order parameter, i.e.
even in time, for both the spin singlet $d_{0\textbf{k}}$ and the
triplet channel
$\textbf{d}_{\textbf{k}}=\{d_{1\textbf{k}},d_{2\textbf{k}},d_{3\textbf{k}}\}$
expressed through the following matrix
$\hat{\Delta}_{\textbf{k}}$:
\begin{center}
$\hat{\Delta}_{\textbf{k}}=
\begin{pmatrix}
-d_{1\textbf{k}} +i d_{2\textbf{k}}  & d_{3\textbf{k}} + d_{0\textbf{k}} \\
  d_{3\textbf{k}} - d_{0\textbf{k}} & d_{1\textbf{k}} +i d_{2\textbf{k}}
\end{pmatrix}
$
\end{center}

\noindent where $d_{0\textbf{k}}=d_{0-\textbf{k}}$ and
$\textbf{d}_{\textbf{k}}=-\textbf{d}_{-\textbf{k}}$
\cite{manfred,Hlubina}. For the classification purposes it is
useful to introduce the vector
$\textbf{q}_{\textbf{k}}=\textbf{r}_{\textbf{k}}+\textbf{p}_{\textbf{k}}$,
with
$\textbf{r}_{\textbf{k}}=d_{0\textbf{k}}\,\textbf{d}_{\textbf{k}}^*+{d_{0\textbf{k}}}^*\,\textbf{d}_{\textbf{k}}$
and $\textbf{p}_{\textbf{k}}=i \textbf{d}_{\textbf{k}}\times
\textbf{d}_{\textbf{k}}^*$, that, for a non-vanishing amplitude,
yields a non-unitary superconducting state with two non equivalent
branches in the excitation spectrum. This is a direct consequence
of the product
$\hat{\Delta}_{\textbf{k}}\,\hat{\Delta}^{\dagger}_{\textbf{k}}$
being not proportional to the identity matrix. The
$\textbf{q}$-vector is decomposed in two components, one lying in
the plane defined by the vectors
$[\textbf{d}_{\textbf{k}},{\textbf{d}_{\textbf{k}}}^*]$ and the
other perpendicular to it. The in-plane term identifies a
non-unitary state with pure mixed singlet-triplet character, while
the perpendicular component is proportional to the magnetic moment
of the Cooper pairs.

To describe such type of superconducting states in systems that
can also lack the inversion symmetry one may use the following
generalized BCS model:

\begin{eqnarray}
H=&&\sum_{\textbf{k},s,s'} ( \xi_{\textbf{k}}
\sigma_{0}+g_{\textbf{k}} \cdot {\bf{\sigma}}_{s,s'}+ {\bf{h}}
\cdot {\bf{\sigma}}_{s,s'}) c^{\dagger}_{{\textbf{k}}s}
c_{{\textbf{k}}s'} +\nonumber \\&& \sum_{{\textbf{k}},s s'}
(\Delta_{{\textbf{k}},s\,s'}
c_{\textbf{k}s}c_{-\textbf{k}s'}+\Delta^{*}_{\textbf{k},s\,s'}
c^{\dagger}_{-\textbf{k}s}c^{\dagger}_{\textbf{k}s'} ) \label{eq1}
\end{eqnarray}

\noindent where ${\bf{\sigma}_0}$ is the identity matrix,
$\bf{\sigma}$ the Pauli matrices,
$\xi_{\textbf{k}}=\varepsilon_{\textbf{k}}-\mu$ is the kinetic
energy measured with respect to the chemical potential $\mu$,
$\Delta_{{\textbf{k}},{s s'}}$ are the equal time components of
the order parameter and $\textbf{h}$ is the magnetic field,
respectively. $g_{\textbf{k}}$ is the antisymmetric spin-orbit
term due to Rashba interaction associated with the breaking of the
parity symmetry and such that $g_{\textbf{k}}=-g_{-{\textbf{k}}}$
due to time-reversal invariance. For the quantitative analysis we
assume a specific tight-binding spectrum in two dimensions of the
form $\varepsilon_{ \textbf{k}}=-4\,[\cos(k_x)+\cos(k_y)]$, in
units of the hopping amplitude, though the results do not depend
on the details of the free electron dispersion and on the
dimensionality.

To search for OTES we explicitly determine for the model
Hamiltonian in Eq. \ref{eq1} the odd-in-time local correlator for
equal-spin pairing that reads:
\begin{eqnarray}
F^{\sigma_i}(t) = \frac{1}{V}\sum_{\textbf{k} } \langle T_{t}
(c_{\textbf{k} \sigma_i}(t)c_{- \textbf{k} \sigma_i}(0)\rangle
\end{eqnarray}
where $V$ is the volume system, $T_{t}$ is the time ordering
operator, $\langle\hat{...}\rangle$ the thermal average and
$\sigma_i$ = $\uparrow$,$\downarrow$ along the direction $i=x,y,z$
in the spin space, respectively.

\begin{table*}
\caption{Cases with non-trivial odd-in-time equal-spin triplet
pair correlations. Y and N indicates non-zero and null amplitude
of the corresponding variable, respectively. $d_0$ and
$\textbf{d}_{\textbf{k}}$ indicate the spin singlet and triplet
part of the order parameter. $\textbf{r}_{\textbf{k}}$ and
$\textbf{g}_{\textbf{k}}$ are the vectors associated with the
non-unitary parity mixing component of the
$\textbf{q}_{\textbf{k}}$-vector and the Rashba interaction in
non-centrosymmetric system, respectively. $\textbf{h}$ is the
applied field. $F^{\mathbf{\sigma}}$ is the odd-in-time correlator
for pairs with equal-spin $\sigma$ along a given direction.
$\angle \textbf{g}$ indicates the directions coplanar to the
$\textbf{g}_{\textbf{k}}$-vector.}
\begin{center}
\begin{tabular}{c||c c c||c c||c c c c}
\hline
\multicolumn{1}{c}{Inversion} & \multicolumn{3}{c}{Magnetic field}
& \multicolumn{2}{c} {Order
parameter} & \multicolumn{4}{c}{OTES correlator} \\
\hline {I} \ & $\textbf{h}$ \ &
$\textbf{h}\cdot\textbf{r}_{\textbf{k}}$ \ &
$\textbf{h}\cdot\textbf{g}_{\textbf{k}}$ \ &  $d_{0\textbf{k}}$ &
$\textbf{d}_{\textbf{k}}$ & $F^{\mathbf{\sigma} || \textbf{h}} $ &
$F^{\mathbf{\sigma} \perp \textbf{h}}$ & $F^{\mathbf{\sigma} \angle \textbf{g}}$ & $F^{\mathbf{\sigma} \perp \textbf{g}}$\\
\hline \hline Y  \ & Y  \ & N \ & N & Y & N & N & Y & N & N  \\
\hline Y \ & Y \ & N \ & N \ & Y & Y & N & Y & N & N \\
\hline Y \ & Y \ & Y \ & N \ & Y & Y & Y & Y & N & N \\
\hline
\hline N \ & N \ & N \ & N \ & Y & Y & N & N & Y & N \\
\hline N \ & Y \ & N \ & N \ & Y & Y & N & Y & Y & N \\
\hline N \ & Y \ & N \ & Y \ & Y & Y & Y & Y & Y & Y \\
\end{tabular}
\end{center}
\label{Tab1}
\end{table*}

The results obtained are reported in the Table \ref{Tab1} where
the conditions and the directions of non zero OTES are indicated.
It is worth pointing out that even-in-time pair correlations are
also present for all the cases associated with equal time
non-trivial order parameters and they can be as well induced by an
applied field.

Let us firstly consider the cases of centrosymmetric
superconductors. In order to understand how odd-frequency pair
correlations occur in parity mixing it is useful to start with the
singlet and triplet states separately. For the pure spin singlet
superconductor the OTES correlations occur only in the presence of
an applied field $\textbf{h}$ and with spin projections in the
plane perpendicular to it. This can be understood considering that
the effect of the field along a given direction, i.e. $z$, is to
induce odd-in-time correlations in the $S_z=0$ triplet channel of
the Cooper pairs \cite{Bergeret2001,Eschrig2008}. Then, due to the
$SU(2)$ spin symmetry, they lead to non-zero amplitude for the
equal-spin correlators at a generic angle $\theta$ in the plane
perpendicular to $\textbf{h}$. Due to the symmetry of the ground
and the excited states the correlation function is symmetric for
$Z_2$ inversion along a given direction in the spin space (see
Fig. \ref{fig:fig1}a). On the other hand, if we consider the case
of a pure spin triplet superconductor, though the spectral
amplitude of the time dependent correlator is non-trivial for a
given $\textbf{k}$-point, due to its odd in space parity the
average over the Brillouin zone leads to a cancellation of the
odd-in-time pair correlations. This result does not depend on the
unitary character of the triplet superconductor.

Taking into account the previous analysis, one can directly
construct the results for centrosymmetric mixed parity
superconductors. There are two relevant cases to consider with
respect to the unitary nature of the ground state, corresponding
to a non-zero amplitude of the vectors $\textbf{r}$ and
$\textbf{p}$. If we have a unitary superconductor, OTES
correlations occur in the presence of an applied field and, as for
the case of the pure singlet, they are $Z_2$ symmetric and in the
plane perpendicular to $\textbf{h}$ (see Fig.\ref{fig:fig1}c).
More interesting is the case of a non-unitary mixed parity state.
For such configurations the main role is played by a non-zero
amplitude of the vector $\textbf{r}$. This vector is non-trivial
if time-reversal is broken for either the triplet or the singlet
component. Indeed, for the spin triplet it can lack time reversal
invariance either in the orbital or in the spin channel. The
latter would imply also a non-zero amplitude for the $\textbf{p}$
component. Alternatively, non-unitary $\textbf{r}$ states can also
be related to a non-trivial phase relation between the singlet and
the triplet order parameter that cannot be eliminated via a global
gauge transformation. For these conditions, as schematically
depicted in Fig. \ref{fig:fig1}d, if the applied field is not
perpendicular to $\textbf{r}$ the OTES occur in all directions
(parallel and perpendicular to $\textbf{h}$) and they also break
the $Z_2$ symmetry along a given spin projection axis (Fig.
\ref{fig:fig1}b). Depending on the material system and on the
$\textbf{k}$-symmetry of the order parameter, if it is possible to
choose a magnetic field such as $\textbf{h} \perp \textbf{r}_k$
for all $\textbf{k}$ then there will be a field direction for
which OTES will lie in a plane otherwise they have a finite
component for any space direction. It is worth pointing out that
the angular dependence of OTES correlations versus $\textbf{h}$
can be used to establish whether mixed parity states exhibit time
reversal symmetry breaking and get insight about the
$\textbf{k}$-structure of the gap amplitude. A possible
application is provided by the system
Pr(Os$_{1-x}$Ru$_x$)$_4$Sb$_{12}$ where a mixed parity phase that
violates time reversal symmetry \cite{Mukuda2010} has been
proposed to interpolate between the singlet and the triplet
configurations achieved at the points $x=0$ and $x=1$,
respectively. As a general remark, the observation of OTES can be
an alternative method to muon spin rotation experiments for the
detection of time reversal symmetry breaking states in mixed
parity configuration.

Let us consider the case of noncentrosymmetric superconductors,
thus mixed-parity states in the presence of an antisymmetric
Rashba type interaction. In this respect, as pointed out in Ref.
\cite{Frigeri2004}, the most convenient configuration for the
triplet $\textbf{d}_{\textbf{k}}$ vector is such that
$\textbf{d}_{\textbf{k}}||\textbf{g}_{\textbf{k}}$, though a
possible misalignment between the two vectors is not completely
prohibited. The analysis has been performed in all the possible
configurations \cite{Gentile2011} while the present results, due
to the limited space, refer only to the physical relevant ones. In
Fig.\ref{fig:fig1}e, we have sketched the zero-field distribution
of OTES correlations in the absence of an external source of time
reversal symmetry breaking. In this case, OTES are coplanar to
$\textbf{g}_{\textbf{k}}$ and exhibit a Z$_2$ symmetric behaviour
at any given spin direction. The application of a magnetic field
modifies the distribution of OTES pair correlations in the spin
space in a way that depends on the amplitude of the product
$\textbf{h}\cdot \textbf{g}$. Only when the field has a parallel
component with respect to $\textbf{g}$ then OTES pair correlations
occur both along the direction of $\textbf{h}$ and perpendicular
to it. A similar behaviour can also happen if the
noncentrosymmetric system is non-unitary (see Fig.
\ref{fig:fig1}f). In this case, the non-orthogonality of the field
with respect to $\textbf{r}$ also leads to OTES pair correlations
in all the spin directions.

A relevant aspect for getting OTES in all spin directions is
related to the structure of the excitations spectra. Indeed, the
spectra at each {\textbf{k}}-point in the Brillouin zone can have
four eigenvalues that fulfill or lack the mirror symmetry when
turning $E$ into $-E$. The most general conditions for which
$H_{\textbf{k}}$, i.e. the $\textbf{k}$-component of the $H$
matrix, has two couples of opposite eigenvalues are given by a
vanishing value for the linear and the cubic coefficients of the
characteristic polynomial $P_k= Det[H_{\textbf{k}}-\lambda
\widehat{1}_4]=a_{0 \textbf{k}} + a_{1 \textbf{k}} \lambda +a_{2
\textbf{k}} \lambda^2 + a_{3 \textbf{k}} \lambda^3 + a_{4
\textbf{k}} \lambda^4$, with $a_{3 \textbf{k}} = -Tr{H_
{\textbf{k}}}= 0$ and $a_{1 \textbf{k}} = -4 (\textbf{h}\cdot
\textbf{r}_{\textbf{k}}+2 \xi_k\textbf{h}\cdot
\textbf{g}_{\textbf{k}}+\textbf{p}_{\textbf{k}} \cdot
\textbf{g}_{\textbf{k}}$). If $a_{3 \textbf{k}}$ and $a_{1
\textbf{k}}$ are both zero, one can exactly demonstrate, by means
of symmetry arguments only, that the longitudinal component of
$F^{\sigma||\textbf{h}}(t)$ ($F^{\sigma\perp\textbf{g}}(t)$) with
respect to the field ($\textbf{g}$-vector) is identically zero
\cite{Gentile2011}.

It is thus the removal of mirror symmetry in the energy spectrum
together with the parity of the OTES $\textbf{k}$-component to
yield missing cancellation for the longitudinal part with respect
to the field when averaging in momentum space.

{\it Conclusions} We have shown that OTES pair correlations can
occur in superconductors with parity mixing. For centrosymmetric
systems this requires the presence of an applied magnetic field
and can lead to an asymmetric distribution for opposite spin
projections depending on the field orientation with respect to the
mixed parity non-unitary component. In materials that lack
inversion symmetry, OTES can occur in zero field with components
coplanar to the $\textbf{g}_{\textbf{k}}$-vector or along all spin
directions when a magnetic field not orthogonal to the
$\textbf{g}_{\textbf{k}}$-vector is applied. Due to the intrinsic
differences of the OTES correlations depending on the magnetic
field direction, it is possible to distinguish between various
types of mixed parity superconducting states. We find that the
removal of quasi-particle degeneracy under parity and time
symmetry is a key aspect for avoiding cancellation in the momentum
space and get OTES pair correlations in all spin directions.

We point out that a possible way to detect OTES is by setting up
an experiment where spin-polarized long range Josephson current is
observed across a diffusive half-metallic ferromagnet at different
uniform orientations of the magnetization. For the cases discussed
in the paper there is no need to have an inhomogeneous profile of
the magnetic pattern or spin active interfaces. The direction of
the magnetization is used as an effective field to generate OTES
along a given direction and at the same time as a filter to select
the OTES pair correlations along a given spin direction. In
particular for the case where OTES are longitudinal to the applied
field it is possible to have spin-polarized pairs crossing the
ferromagnet without the need of inhomogeneous profile in the
magnetization.

This result can lead to a significant advance in the context of
hybrid systems where the occurrence of mixed parity and
subdominant components close to the interface can be the sources
for time and parity symmetry breaking in the energy spectra and in
turn for equal-spin odd-frequency pair correlations even without
magnetic inhomogeneities. This observation can be used to engineer
solutions where long range pairing currents can be induced by
including sources of parity mixing at the interface between
singlet superconductors and ferromagnets, especially in the form
of Rashba type junctions, and by properly selecting the
magnetization direction. Since this parity mixing mechanism works
for a uniform magnetic profile it can also account for the
persistence of the long range supercurrents in hybrids even when
the applied magnetic field is such that no magnetic domains are
present in the ferromagnet.

\begin{acknowledgments}
We acknowledge useful discussions with I. Vekhter and J. Aarts.
The research leading to these results has received funding from
the FP7/2007-2013 under the grant agreement No.\ 264098 - MAMA.
\end{acknowledgments}



\end{document}